\documentclass[manuscript,screen]{acmart}

\usepackage{lscape}
\usepackage{csquotes}

\AtBeginDocument{%
  \providecommand\BibTeX{{%
    \normalfont B\kern-0.5em{\scshape i\kern-0.25em b}\kern-0.8em\TeX}}}

\setcopyright{acmlicensed}
\acmJournal{JOCCH}
\acmYear{2023} \acmVolume{1} \acmNumber{1} \acmArticle{1} \acmMonth{1} \acmPrice{15.00}\acmDOI{10.1145/3605910}

\begin{document}

\title{Moving from ISAD(G) to a CIDOC CRM-based Linked Data Model in the Portuguese Archives}

\author{Inês Koch}
\email{ines.koch@inesctec.pt}
\orcid{0000-0002-9363-9713}

\affiliation{%
  \institution{Faculty of Engineering of the University of Porto and INESC TEC}
  \city{Porto}
  \country{Portugal}
}

\author{Carla Teixeira Lopes}
\orcid{0000-0002-4202-791X}
\affiliation{%
  \institution{Faculty of Engineering of the University of Porto and INESC TEC}
  \city{Porto}
  \country{Portugal}
}
\email{ctl@fe.up.pt}

\author{Cristina Ribeiro}
\orcid{0000-0001-6150-0090}
\affiliation{%
  \institution{Faculty of Engineering of the University of Porto and INESC TEC}
  \city{Porto}
  \country{Portugal}
}
\email{mcr@fe.up.pt}

\renewcommand{\shortauthors}{Koch, Teixeira Lopes, Ribeiro.}

\begin{abstract}
Archives are facing numerous challenges. On the one hand, archival assets are evolving to encompass digitized documents and increasing quantities of born-digital information in diverse formats. On the other hand, the audience is changing along with how it wishes to access archival material. Moreover, the interoperability requirements of cultural heritage repositories are growing. 
In this context, the Portuguese Archives started an ambitious program aiming to evolve its data model, migrate existing records, and build a new archival management system appropriate to both archival tasks and public access. The overall goal is to have a fine-grained and flexible description, more machine-actionable than the current one. 
This work describes ArchOnto, a linked open data model for archives, and rules for its automatic population from existing records. ArchOnto adopts a semantic web approach and encompasses the CIDOC Conceptual Reference Model and additional ontologies, envisioning interoperability with datasets curated by multiple communities of practice. Existing ISAD(G)-conforming descriptions are being migrated to the new model using the direct mappings provided here. 
We used a sample of 25 records associated with different description levels to validate the completeness and conformity of ArchOnto to existing data.
This work is in progress and is original in several respects: (1) it is one of the first approaches to use CIDOC CRM in the context of archives, identifying problems and questions that emerged during the process and pinpointing possible solutions; 
(2) it addresses the balance in the model between the migration of existing records and the construction of new ones by archive professionals; and (3) it adopts an open world view on linking archival data to global information sources.
\end{abstract}

\begin{center}
\fbox{\parbox{\textwidth}{
%\vspace{0,1cm}
\textcopyright{} Inês Koch, Carla Teixeira Lopes and Cristina Ribeiro, 2023. This is the author's version of the work. It is posted here for your personal use. Not for redistribution. The definitive Version of Record was published in ACM journals - Journal on Computing and Cultural Heritage, \url{http://dx.doi.org/10.1145/3605910}.
}}
\end{center}


\setcopyright{acmlicensed}
\acmJournal{JOCCH}
\acmYear{2023} \acmVolume{1} \acmNumber{1} \acmArticle{1} \acmMonth{1} \acmPrice{15.00}\acmDOI{10.1145/3605910}


\maketitle

\section{Introduction}
National archives are stable institutions by definition. They have to ensure that their infrastructures and processes can be trusted with our shared memory for the long term, measured in centuries. Their mission is centered on preservation: making sure archival assets are kept in the right conditions, promoting recovery actions for those that become fragile, and keeping the whole in the best conditions that the available staff and funding allow. They are, therefore, cautious regarding the adoption of disruptive solutions. On the other hand, they are confronted with an ever-changing panorama of media, digital supports (requiring specific technology to access the contents), and even target audience.

Digital assets are challenging archives in several aspects. They require quite different infrastructures to be maintained and are not tolerant of so-called ``benign neglect'': unlike traditional materials that degrade continuously and over long periods, neglect in digital entails inexorable loss. Maintaining digital assets has required archives to be equipped with vast and safe digital infrastructures, with the corresponding staff. Also, the transition to digital creates expectations concerning the digitization of analog supports, a challenging task given the volumes of information available, and the brittleness of the treasures kept in the archives.
Investments in digital infrastructure are also hard to plan in the long term, given the pace of change of digital technologies.

Despite their focus on preservation, archives are more and more responding to findability and access requirements, namely for laypeople, in quite a shift from their focus on scholars and specialized access. These requirements have led to a more systematic view of the archival contents and their organization according to standards for identification and description. The General International Standard Archival Description, ISAD(G), has been designed to deal with extensive collections of heterogeneous objects and adopts a uniform description to make them manageable. Description can be inherited from a fonds down to a document series or a single document, making description in large collections feasible and accommodating uneven description efforts that take into account the potential for access and reuse.

Access is not the only requirement when people explore an extensive collection, such as the public archives. The information therein also needs to be linked to other sources. For example, the biographies of the people involved may complement the record for an international treaty, and the toponymic reference in a medieval document may connect to its present-day location.

In this context, we present the joint effort of a team from Information and Computer Science and the archival experts from the Portuguese National Archives. The EPISA project (Entity and Property Inference for Semantic Archives) is rethinking the existing archival system, assisting in the transition from the ISAD(G) multilevel description model to a linked data model used to represent archival information on a prototype open-source knowledge platform. Current descriptions will be used by Artificial Intelligence algorithms to extract entities and infer relationships between those entities. The prototype knowledge graph and corresponding user interfaces will support applications for specialists and also for the public.

This work describes ArchOnto, a linked open data model for archives, and the rules for its automatic population from existing records. ArchOnto encompasses the International Committee for Documentation Conceptual Reference Model (CIDOC CRM)  of the International Council of Museums (ICOM) and additional ontologies, envisioning interoperability with external sources to build a more robust user experience. As this is one of the first approaches to use CIDOC CRM in the context of archives, several problems and questions emerged during the process that we describe here, and solutions are proposed.

This article is organized as follows. Section~\ref{sec:SOTA} has a brief description of the Portuguese National Archives, presents and compares the main description standards in cultural heritage, reports on existing applications of CIDOC CRM in the context of museums and archaeology, and describes works focused on migration between description standards. In Section~\ref{sec:ArchOnto} we detail ArchOnto and its development. To validate the appropriateness of ArchOnto with existing data, we manually represented a sample of 25 records, an experiment described in Section~\ref{sec:ManualRepresentation}. Based on this experiment, we produced a set of rules, detailed in Section~\ref{sec:migration}, that are currently being used in the first stage of the migration process. We finish with a discussion of results in Section~\ref{sec:Discussion} and present our main conclusions in Section~\ref{sec:Conclusions}.

\section{Background and State of the Art}
\label{sec:SOTA}

Cultural heritage institutions have pioneered the definition of standards for resource description, in line with the missions of libraries, archives, and museums. They evolved as specialized, dedicated resources and have been to some extent adopted in the more flexible web-age models for resource description~\cite{Weibel2000}. 
The Portuguese National Archives have long adopted the ISAD standards and built a dedicated archival information system to support their operation, including interfaces for archivists and for the web, while keeping record of conservation activities and the management of the physical location of assets and the reading rooms.

\subsection{Portuguese Archives}
The General Directorate for Book, Archives, and Libraries (DGLAB) is a public body under the responsibility of the Portuguese Ministry of Culture. This structure comprises two national archives (Torre do Tombo National Archive - ANTT - and the Portuguese Center of Photography), the Overseas Historical Archive, and 16 regional archives at the district level \cite{RedeDGLAB}. Torre do Tombo is one of the oldest national archives in the world and one of the oldest institutions in Portugal. The ANTT curates a unique collection of historical and contemporary objects that were accumulated since the 9th century.

DigitArq \cite{DBLP:conf/elpub/RamalhoF04} is a custom platform commissioned by the National Archives and is currently used for archival description by the DGLAB. DigitArq also provides public access to online services and is available at \url{https://digitarq.arquivos.pt/}. Remotely, users can access the archives, look at scanned documents, request digital reproductions, book documents for in loco examination, or request certificates.

DGLAB has about 2.8 million records, most of them under the custody of ANTT.
The oldest records were produced in the 9th century, and the number of records has been steadily growing over time. Fonds are organized in groups, such as Central and Local Administration; Collections; Companies; Judicial; Monastics; Notaries; Parish; and Personal. Collections include records from \textit{Estado Novo}, Portugal's Ancien Régime; records from contemporary, ecclesiastical, monastic and conventual institutions; records of archives of individuals, families, associations, companies, commissions and congresses; and records of photographic archives.

Most of the records (about 1.4 million) are in Portuguese, and the second language is Spanish, with 30.8 thousand records. As expected, the number of records with a digital representation has been growing in the last decade. The main format used for these is tiff, followed by jpg and pdf. As expected, the most common levels of description are the `file', corresponding to compound documents in DigitArq, with 1.75 million records, and `item', used for simple documents, with 0.8 million records. Regarding the elements of description, the ones with a large median for the number of characters are the `Biographical History', followed by the `Finding Aids', `Related Units of Description' and `Scope and Content'. 

\subsection{Standards for Description in Cultural Heritage}
The International Council on Archives (ICA) is responsible for the development of the ISAD(G) General International Standard Archival Description \cite{InternationalCouncilonArchives2016}, a widely adopted standard. It is characterized by a multilevel structure that allows description to proceed from general to specific, representing each fonds' context, hierarchical structure and components. The second edition of the standard dates from 2000 and includes general guidelines for the preparation of archival descriptions. ISAD(G) was the basis for the development of the Encoded Archival Description (EAD), a standard maintained by the Society of American Archivists and the Library of Congress~\cite{pittiEAD}. The EAD is an XML language that is used as an interoperability layer between the archival standards and the technological platforms supporting them. 

More recently, ICA started developing a new data model that builds on the concept of linked open data --- RiC-CM. This new model was developed considering data models developed by archival communities in several countries, namely Australia, Spain and Finland. 

The Australian Government Recordkeeping Metadata Standard (AGRkMS) includes information about records and the contexts where they are captured and used~\cite{Australia2015}. The latest version (2.2) was issued in 2005 to revise the first version, dating from 1999.  
This standard includes the minimum metadata necessary to ensure that records remain accessible and usable over time. It also comprises the metadata needed to manage the preservation of digital records for ongoing agency business needs and metadata for records held in a digital archive~\cite{Australia2015}.

In Spain, the \textit{Comisión de Normas Españolas de Descripción Archivística} (CNEDA) has developed \textit{Modelo Conceptual de Descripción Archivística y Requisitos de Datos Básicos de las Descripciones de Documentos de Archivo, Agentes y Funciones (MCDA)} -- Conceptual Archival Description Model and Basic Data Requirements for the Descriptions of Archival Records, Agents and Functions. 
In this conceptual model of archival description, the types of entities (record, agent, role, standard, concept, object or event, location) are presented, as well as the relationships between entities of these types, and the attributes of three types of entities (record, agent, role)~\cite{ComisionNormasEspanolasdeDescripcionArchivistica2012}.

In Finland, the Finnish Conceptual Model for Archival Description,  presented from here on as FCMAD, was developed for archival description and consists of an ontological conceptual model and metadata models, implementing the conceptual model in practice.
The model, developed by the National Archives of Finland (\textit{Arkistolaitos}), advocates separate models for the main description entities, along with the relations that hold between them. It recognizes a set of entities: function, agent, manifestation, expression, item, information resource, life cycle, mandate (norm), place, time, and subject~\cite{Llanes-Padron2017}.

The goal of the Records in Contexts Conceptual Model (RiC-CM)~\cite{InternationalCouncilonArchives2016} is to reconcile, integrate, and build on four previous ICA standards: ISAD(G); the International Standard Archival Authority Records --- Corporate Bodies, Persons, and Families (ISAAR(CPF)); the International Standard Description of Functions (ISDF); and the International Standard Description of Institutions with Archival Holdings (ISDIAH). In July 2021, ICA released a full version of the 0.2 version of this model and, in February 2021, the 0.2 version of the Records in Contexts Ontology.

The CIDOC Conceptual Reference Model (CIDOC CRM)~\cite{ICOM/CIDOCCRMSpecialInterestGroup2019} 
%, a formal ontology, 
was developed in the scope of museums by the International Committee for Documentation (CIDOC) of the International Council of Museums (ICOM). This model, which aims to exchange, mediate, and integrate heterogeneous sources of information related to cultural heritage, is under active development by the CIDOC CRM Special Interest Group (CIDOC CRM SIG). The CIDOC CRM went through several changes over the years, and it became an ISO standard in 2006 (ISO 21127), with a revision in 2014 (ISO 21127: 2014). At present, CIDOC CRM is the only ontology in the Cultural Heritage domain to have this official status, recognizing its acceptance in the community and contributing to reinforce it~\cite{Bruseker2017}. It has a strong emphasis on events and is also quite detailed concerning the representation of people, places, and time periods, concepts that are quite central in archival description.

In libraries, the International Federation of Library Associations and Institutions (IFLA) developed the Functional Requirements for Bibliographic Records (FRBR), with two main goals. The first is to provide a clearly defined structure for relating the data in the bibliographic records to the users' needs. The second is to recommend a basic level of functionality for records created by national bibliographic institutions~\cite{IFLAStudyGroup2009}.

Table \ref{tab:standardcomparison} provides a high-level view of some features of ISAD(G), RiC-CM, CIDOC CRM, AGRkMS, MCDA, and FCMAD. We exclude FRBR due to its smaller expression in the context of archives. The ICA, Australian, Spanish, and Finish models consider the hierarchical structure intrinsic to the archives. CIDOC CRM, RiC-CM, AGRkMS, MCDA, and FCMAD comply with semantic web principles and, therefore, aim to represent cultural heritage data as linked data. While the ISAD(G) standard has a limited number of elements, some of which have values that can be quite complex, the more recent models have a number of properties an order of magnitude larger, attesting to their more atomized representation of knowledge. All models have institutional supervision in the corresponding working groups under well-established cultural heritage institutions. While ISAD(G) has been supported in various archival information systems over the years, the RiC-CM still lacks the test of actual deployment. Concerning the models developed by national institutions, these have already been applied in the corresponding archives. 

\begin{table}
  \caption{Standards Comparison}
  \label{tab:standardcomparison}
  \begin{tabular}{lllllll}
    \toprule
    &ISAD(G)& RiC-CM & CIDOC CRM & AGRkMS& MCDA & FCMAD\\
    \midrule
    Hierarchical & \checkmark  & \checkmark & x &\checkmark&\checkmark&\checkmark\\
    LOD & x & \checkmark & \checkmark & \checkmark &\checkmark &\checkmark \\
    Ontology & x & \checkmark & \checkmark& N/A & N/A &\checkmark\\
    Number of properties & x & 449  & 285  & 70 & N/A &16 \\
    Number of elements & 26 & x & x & x & x & x\\
    Supervising institution & ICA & ICA & ICOM& NAA&CNEDA& Arkistolaitos\\
    Years of institutional supervision & +25 years & +5 years & +25 years& +15 years&+10 years& +5 years\\
    Implementation on archives & Custom & None & (none known)&Australia&Spain&Finland\\
    Most recent version & 2000 & 2021&2021 & 2015 & 2012 &2013\\
    Support group & ICA & ICA(EGAD) & CIDOC CRM SIG & NAA & CNEDA & Arkistolaitos\\
  \bottomrule
  \multicolumn{7}{c}{\small N/A -- Not Publicly Available; NAA -- National Archives of Australia}
\end{tabular} 
\end{table}

\subsection{Applications of CIDOC CRM}
Due to its maturity, CIDOC CRM has already been adopted in several areas of knowledge. In the scope of museums, the model was adopted by the Museo del Prado~\cite{Prado2016} and the British Museum~\cite{Bruseker2017}. In addition to museums, the model was used in archaeology, namely, in the Ariadne Project~\cite{ARIADNE2014}, in music through the DOREMUS Project~\cite{DOREMUS} and EthnoMuse~\cite{EthnoMuse}, and in archives through the ICON~\cite{Almeida2018} and EPISA~\cite{ 10.1007/978-3-030-54956-5_10} projects.

In Madrid, Spain, the Museo del Prado developed a knowledge graph that is based on a broad group of ontologies from several domains, integrating them into a shared framework or ontological narrative representing the overall group of activities taking place in the museum setting~\cite{Prado2016}. Among these activities are the documentation associated with conservation processes, the communication with the media, and the online sale of objects from the Prado store. The main entities in the Prado semantic network, namely Work of Art, Author, Exhibition, and Activity~\cite{10.1007/978-3-319-70863-8_31}, are represented according to the CIDOC CRM standard. The FRBR ontology was adopted to represent bibliographic records. The Museum has identified other concepts relevant to its activity, namely those concerning news, the regulations for contracts in the public sector, and human resources. For each of these areas, existing ontologies were adopted and brought together with CIDOC CRM and FRBR in the Digital Semantic Model for the museum\footnote{Website Museo del Prado - Digital Semantic Model -- last consulted on 10/01/2022 -- Available at \url{https://www.museodelprado.es/modelo-semantico-digital/modelo-ontologico}}.

In the British Museum, the first arts organization in the United Kingdom to publish its collection semantically, CIDOC CRM is used to establish links with other institutions\footnote{Website The National Archives -- last consulted on 10/01/2022 -- Available at: \url{https://webarchive.nationalarchives.gov.uk/20170803104533/http://www.britishmuseum.org/about_us/news_and_press/press_releases/2011/semantic_web_endpoint.aspx}}. 

The purpose of the semantic model at the British Museum is to provide a new degree of accessibility, while allowing others to work closely with the data, obtain new insights and produce innovative applications~\cite{Oldman2014}. 
This semantic model was developed in the scope of the Research Space Project, a project that has built the integrative data infrastructure for the web presence of the collections of the British Museum~\cite{Bruseker2017}.

In archeology, extensions to CIDOC CRM, namely CRMba and CRMarchaeo were  proposed. The former intends to model the complexity of a built structure from the perspective of the archaeology of buildings, while the latter was developed to model the processes involved in investigating subsurface archaeological deposits~\cite{10.1007/s00799-016-0193-3}. 
At the European level, the Advanced Research Infrastructure for Archaeological Dataset Networking in Europe (ARIADNE) project stands out in the area of archeology. The ARIADNE data portal uses the CIDOC CRM to implement archaeological interoperability at a conceptual level. 
This project is also contributing to the development of CRMarcheo, the archaeological extension of CIDOC CRM aimed at deep data integration~\cite{ARIADNE2014}.

Also in the archeology domain, the Arches Project was designed to take advantage of the CIDOC CRM approach to create meaningful data relationships between heritage resources and groups, activities, actors, events, and information media~\cite{ArchesCIDOC2021}. It  is an open-source software platform developed jointly by the Getty Conservation Institute and World Monuments Fund for cultural heritage data management~\cite{Arches2021}.

In music, CIDOC CRM was used in the DOREMUS Project to semantically describe the catalogs of musical works and events of three French institutions --- the French National Library (BnF), Radio France, and the Philharmonie de Paris. 
One significant contribution of this project was the creation of the DOREMUS Ontology, which extends the CIDOC CRM and FRBRoo models to represent bibliographic information, adapting it to the domain of music~\cite{DOREMUS}.

CIDOC CRM and the FRBRoo were also used in the EthnoMuse digital library to represent processes and relations in folk song and music realizations. The EthnoMuse digital library consists of novel tools and techniques that assist with digitizing the Institute of Ethnomusicology collections. These support the ongoing production and post-production processes related to field recordings and documentation of Slovenian folk song music and dance, to enable the manipulation and search of the library’s contents~\cite{EthnoMuse}.

In the archives, CIDOC CRM was used in two interlinked projects, ICON and EPISA. These two projects, developed by a multidisciplinary group, use CIDOC CRM to represent archival records and aim to build a new archival technical data infrastructure suited to archival specialists and to external users~\cite{Almeida2018}. Although the RiC-CM is being developed in the archives, at the beginning of the development of these projects, in 2018, it was still at a preliminary stage, which led to the choice of the CIDOC CRM which was already in a more mature stage of development~\cite{ 10.1007/978-3-030-54956-5_10}.

\subsection{Semantic mapping of description standards to CIDOC CRM}
The increasing demand for global search, comparative studies, and data migration between heterogeneous sources of cultural contents requires the semantic interoperability of CIDOC CRM with other description standards~\cite{Doerr2003}.

A semantic mapping of the EAD to CIDOC CRM is one of the first approaches to such interoperability~\cite{theoDoerr2001}.
More specifically, this maps the EAD Document Type Definition (DTD) elements to CIDOC, showing their representation in the CIDOC CRM ontology both visually and through formal rules. 

A more recent work also focuses on the mapping between EAD and CIDOC CRM~\cite{mapping}. It proposes the semantic organization of an archive in 3 hierarchies: physical, informational, and linguistic. The mapping of EAD elements and attributes follows from this organization and is expressed in the Mapping Description Language (MDL).

The semantic integration of CIDOC CRM with Dublin Core has also been studied~\cite{semanticIntegration2009}. It consists of a path-oriented crosswalk from Dublin Core Collections Application Profile (DCCAP) to CIDOC/CRM ontology and vice versa. 
The conclusion is that the mapping is complex, given the event-based character of CIDOC/CRM that requires the use of intermediate activity and event entities.

\section{ArchOnto}
\label{sec:ArchOnto}

Modeling ontologies through the adoption of existing ontologies, builds on others' work and experience, promotes interoperability and is inline with the ideas promoted by the semantic web. CIDOC CRM is a stable model, accepted as an international standard and with ongoing work with the library community. By the time this work started, the model had already been explored by the specialists at DGLAB, and therefore it provided a common language for the project team.

ArchOnto is the name of the new model supporting the representation of assets at DGLAB. Its goal is to serve as the foundation for the development of a knowledge graph prototype and new applications for the archives. 

\subsection{Development process}
With CIDOC CRM as the basis for the development of the data model for the archives, we did an in-depth study of the standard. We considered the broad set of concepts in CIDOC CRM to represent the information in existing records in DigitArq, which are described with the ISAD(G) standard. 
Many essential concepts in archives are well established in digital humanities, and therefore already present in CIDOC CRM. Others are specific to the domain of the archives and have no direct representation in CIDOC CRM. An example is the \textit{Level of Description}, specific to the hierarchical representation of ISAD(G), a mandatory element in existing records, and one that we intend to maintain in the new model.

As we recognized that CIDOC CRM alone would not provide the expressiveness for all the fundamental concepts in archives, we considered extensions or other ontologies. The ArchOnto data model\footnote{OWL version available on GitHub - ArchOnto2021:
\url{https://github.com/feup-infolab/archontology/tree/master/ArchOnto_2021}} 
emerged. 
At first, Archonto included CIDOC CRM and some extensions, but it soon became clear that we should integrate separate ontologies instead of extensions in some situations. Distinct ontologies make it easy for an ontology created specifically for this model to be replaced by a reference ontology with the same purpose, taking advantage of new developments as they appear.

Beyond the separate ontologies, ArchOnto includes some extensions to CIDOC CRM, such as a set of subclasses of \textit{E55 Type}. This class is the entry point for the representation of categories of objects for which the archives use controlled vocabularies. Without the subclasses, \textit{E55 Type} would be used for such diverse types as an identifier type, taking values such as reference code or physical location, or a name type, instantiated as first name, surname, nickname. The use of \textit{E55 Type} for  the whole range of such types in the archives would not allow additional control on the used vocabularies, and therefore subclasses were created. For each subclass, the instances will be determined by existing vocabularies for the corresponding type. Table \ref{tab:classes_ArchOnto}, in Appendix A, shows that we followed the same logic in the creation of subclasses for \textit{E35 Title}, \textit{E39 Actor}, and \textit{E54 Dimension}.

\subsection{ArchOnto Architecture}
\begin{figure}
\centering
  \includegraphics[scale=0.40]{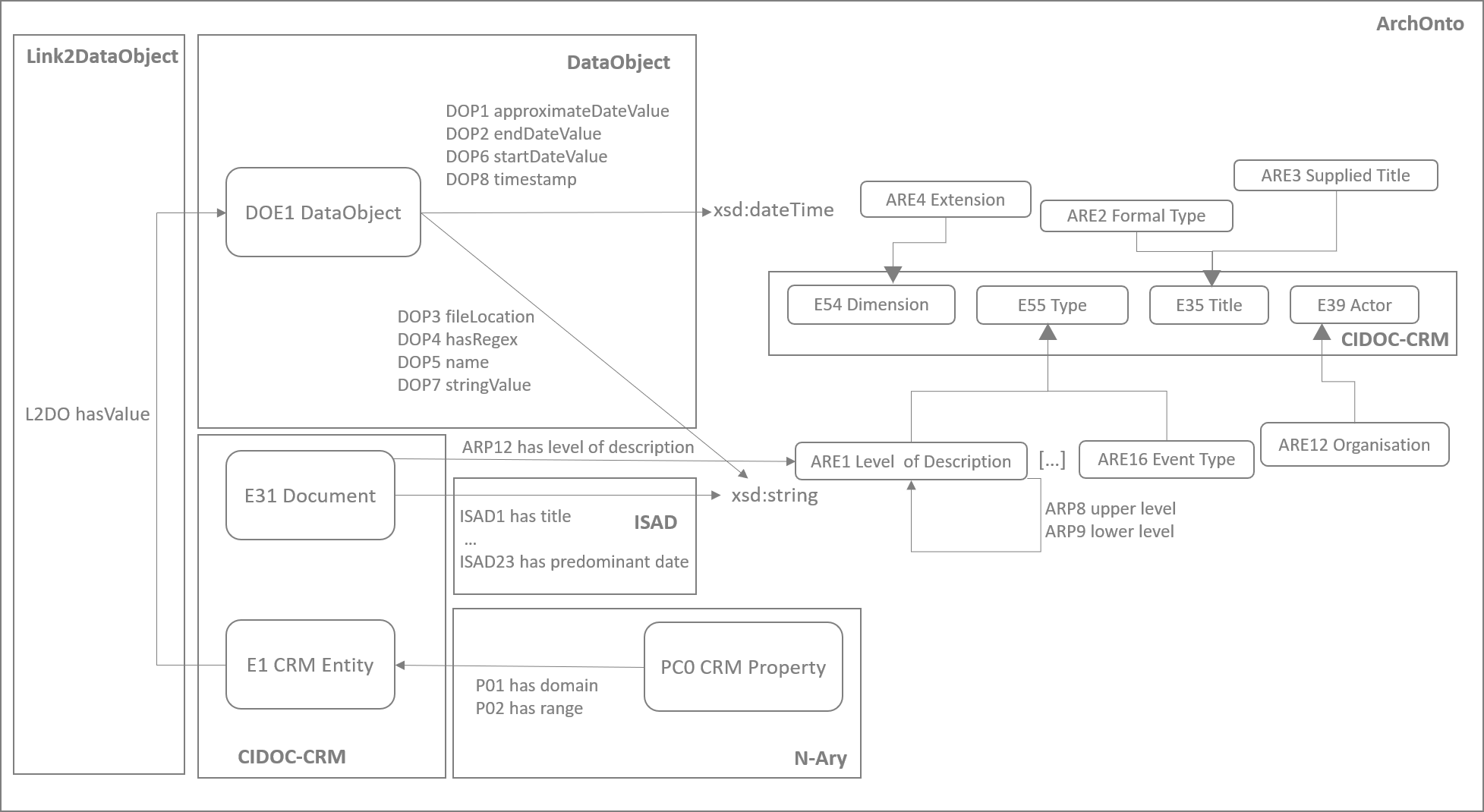}
\caption{ArchOnto architecture. \textit{ARP12 has level of description} is a subproperty of CIDOC CRM \textit{P2 has type} property.}
\label{fig:modelo_ontologia}
%\note{\textit{ARP12 has level of description} is a subproperty of CIDOC CRM \textit{P2 has type} property.}
\end{figure}

ArchOnto is presented as a modular data model that integrates five ontologies, dealing with complementary aspects of the archival domain, as illustrated in Figure~\ref{fig:modelo_ontologia}. The five ontologies are CIDOC CRM (base ontology), N-ary, DataObject, ISAD Ontology, and Link2DataObject. Figure~\ref{fig:modelo_ontologia} also shows ArchOnto's classes and properties and their relation with other ontologies.

\begin{figure}
\centering
  \includegraphics[width=\textwidth]{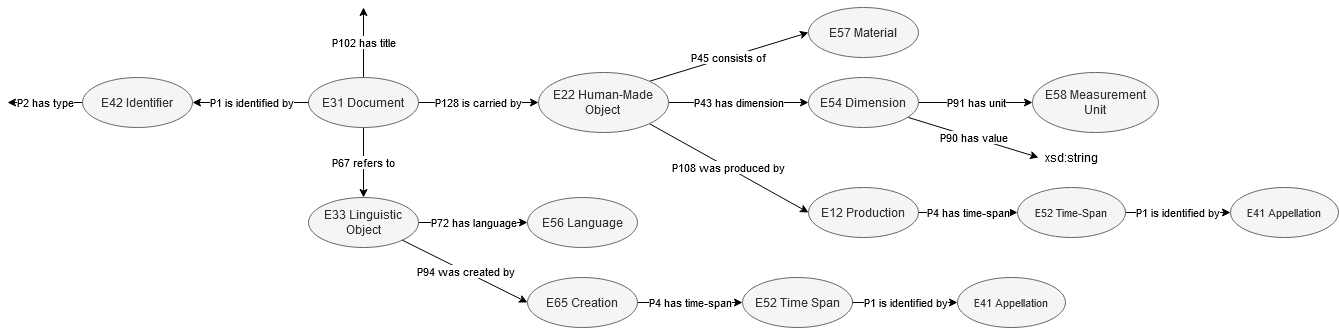}
\caption{Outstanding CIDOC CRM classes and properties in ArchOnto}
\label{fig:archonto-CIDOC2}
\end{figure}

CIDOC CRM is the core of ArchOnto, providing the concepts and properties to capture archival records' essential features, e.g. event, date, location, person, group. Figure~\ref{fig:archonto-CIDOC2} shows the CIDOC CRM classes and properties that are most likely to be used to represent an archival record, that typically has information such as the title, language, document identifiers, and their dimension and support.

The N-ary ontology provides a systematic way to represent non-binary associations, i.e., those that connect more than two individuals. Representing a person's role in a given event is an example of this kind of association because it involves relating an event, a person, and a role. N-ary takes into account the CIDOC CRM recommendations\footnote{\url{http://www.cidoc-crm.org/sites/default/files/Roles.pdf}} for the representation of tuples with arity higher than two. In ArchOnto, this is modeled as a complementary ontology that follows the recommendations of CIDOC. As a separate ontology, it is easier to replace it or even dismiss it in case CIDOC CRM evolves to include these concepts. 
Figure~\ref{fig:modelo_ontologia} illustrates the classes and properties used in the N-ary ontology. Tables \ref{tab:classes_ArchOnto} and \ref{tab:archontoProperties}, in  Appendices A and B, show how the classes and properties from N-ary fit in the ArchOnto's hierarchy.

DataObject is an auxiliary ontology created to deal with literal values and their validation in ArchOnto. The goal is for each individual that has a representation as a simple type, such as a date or a string, to be validated based on the corresponding class. DataObject comprises classes and data properties for the simple types in the ontology. These classes and properties are illustrated in Figure \ref{fig:modelo_ontologia} and more systematically in Appendices A and B. In this ontology, the names of classes are preceded by DOE (from \textbf{D}ata\textbf{O}bject \textbf{E}ntity), and the names of properties are preceded by DOP (from \textbf{D}ata\textbf{O}bject \textbf{P}roperty).

The ISAD Ontology was designed to take into account the archival practice in legacy records and to facilitate the migration of such records. CIDOC CRM, and therefore ArchOnto, has more fine-grained information than the existing records. In parallel with the evolution of the model and the archival applications, the creation of new records by the experts will comply with this increased granularity. 
In the migration process, textual elements from existing records are mined to extract relevant information for populating the new model. This extraction, albeit successful, will generate text fragments that can hardly convey the full meaning of the existing text. 
The ISAD Ontology matches the ISAD(G) standard elements that can be directly obtained from existing records. The information is represented in this ontology with Data Properties and is not atomized (as shown on Table \ref{tab:archontoProperties}, in Appendix B). The ontology can be regarded as an intermediate representation for legacy records, making it possible, whenever necessary, to check the contents of a given element against the information extracted thereof. The ISAD Ontology will only be used for the existing records in DigitArq and not for newly created records based on ArchOnto.

Finally, the Link2DataObject ontology makes the connection between the CIDOC CRM and the DataObject ontologies (Table \ref{tab:archontoProperties}). It has a single property to make this connection --- \emph {L2DO hasValue}. Link2DataObject is separated from DataObject for practical reasons: if another ontology replaces DataObject, the link to the CIDOC CRM is still required.

In addition to these five imported ontologies, ArchOnto has its classes and properties, 
%structure classes and properties 
created as extensions to CIDOC CRM. %Classes are used to specialize concepts already present in CIDOC CRM, but that need to be adapted for the archives. 
Classes are used to specialize concepts already present in CIDOC CRM but which needed to be adapted to include controlled vocabularies that exist in the archives. Figure~\ref{fig:modelo_ontologia} and Table~\ref{tab:classes_ArchOnto} in Appendix~A show the classes that are part of this ontology and how they relate to CIDOC CRM. In addition to these classes, the ontology also includes three properties. \textit{ARP12 has level of description} relates a given document to its level of description. \textit{ARP8 upper level} and \textit{ARP9 lower level} (inverse of each other) relate a level of description with the upper (respectively lower) levels where it may be aggregated (aggregate). This provides a control structure for the admissible nesting of units of description. Names for classes are preceded by ARE (from \textbf{AR}chival \textbf{E}ntity), and those for properties are preceded by ARP (from \textbf{AR}chival \textbf{P}roperty).

To facilitate and validate the population of ArchOnto, we have defined controlled vocabularies that can be used in one or more classes. As an example, we present some of these vocabularies in Table~\ref{tab:vocabularios}, along with the classes and example values.

\begin{table}[ht]
  \caption{Controlled vocabularies}
  \label{tab:vocabularios}
\begin{tabular}{lll}
\toprule
Class & Vocabulary & Example Values \\ 
\midrule
ARE1 Level of Description & Level of description & Fonds; Series; Section; File; Item\\ 
\midrule
\begin{tabular}[c]{@{}l@{}}
ARE2 Formal Title\\ ARE3 Supplied Title\end{tabular} & Title Type & Formal, Supplied \\
\midrule
ARE5 Identifier Type & \begin{tabular}[c]{@{}l@{}}Identifier of collective \\ person/group\end{tabular} & \begin{tabular}[c]{@{}l@{}}PT; VCT; AGH01; 161016; ADLSB; \\ 600084892; PT-LiBN\end{tabular} \\
\midrule
\begin{tabular}[c]{@{}l@{}}ARE6 Date Type \\ ARE9 Date Certainty\end{tabular} & Type of time period & \begin{tabular}[c]{@{}l@{}}Exact dates; Inferred dates; Predominant \\ dates \end{tabular}  \\
\midrule
ARE7 Name Type & \begin{tabular}[c]{@{}l@{}}Type of name of collective\\  person/group\end{tabular} & \begin{tabular}[c]{@{}l@{}}Authorized form of name; Another form of \\ the name; Parallel name form \end{tabular}  \\ 
\midrule
ARE8 Role Type & Role played & Producer; Material Author; Recipient \\
\midrule
ARE11 Documentary Typology & Documentary Typology & Certificate; Income book; Patent \\
\midrule
ARE13 Subject Type & Subject & Education; Science; Law; Management\\
\midrule
ARE14 Place Type & Type of jurisdictional entity & \begin{tabular}[c]{@{}l@{}}Ocean; Archipelago; Mountain range;\\ Country; District \end{tabular}  \\
\midrule
ARE15 Acquisition Type & \begin{tabular}[c]{@{}l@{}}Transfer of Custody /\\ Acquisition Identifier\end{tabular} & \begin{tabular}[c]{@{}l@{}}Purchase; Giving; Donation; Deposit; \\ Swap; Legacy; Reintegration; Transfer \end{tabular} \\
\midrule
ARE16 Event Type & Event Type & Evaluation; Expertise; Financial management\\
\midrule
E56 Language & Language Identifier & Portuguese; Latin; French; Greek\\
\midrule
E57 Material & Support & Paper; Parchment; Photosensitive film\\ 
\midrule
E58 Measurement Unit & Measurement Unit & Centimeter; Gram; Byte; Minute; Pack \\
\midrule
E98 Currency & Currency & Euro; Dollar; Kwanza\\
\bottomrule
\end{tabular}
\end{table}

%%%%%%%%%%%%%%%%%%%%%%%%%%%%%%%%%%%%%%%%%%%%%%%%%%%%%%%%%%%%%%%%%%%%
\section{Manual Representation}
\label{sec:ManualRepresentation}
The Portuguese National Archives hold a vast number of cultural objects. To validate the appropriateness of ArchOnto with existing data, we manually represented a sample of records. In Section~\ref{sec:Sample} we describe this sample and how it was obtained. 

A manual representation process was then performed for the record sample, according to the data model. The representation process ran in an iterative way. The representation of the records was carried out for subsets of five records, and the representation for each subset was discussed with the archival specialists. We manually introduced all individuals and property instances in Protégé, the tool used to develop the ontology. Due to space restrictions, in this paper's context, we only include ArchOnto's representation of one of the records (see Section~\ref{sec:Representation}). The complete set of representations is available as supplemental material~\footnote{Available at: \url{https://github.com/feup-infolab/archontology/tree/master/ArchOnto2021_25records}}. 

\subsection{Sample}
\label{sec:Sample}
Experts of the Portuguese National Archives selected twenty-five records representative of the various description levels and groups of fonds. These records are listed in Table~\ref{tab:sample} and include ten records at different levels of description within the same fonds plus fifteen records of representative groups of fonds. 

In the record sample, some documents are relevant to the Portuguese cultural heritage. In the documents of the \textit{Juízo da Índia e Mina} fonds, for instance, there are records related to the \textit{Ultramar} (former Portuguese colonies), providing evidence of processes related to the commercial routes used, as well as the products that were marketed, the ships and their types. 
In addition, records from representative classes include those relating to monastic, personal, notary, and central administration documents.

\begin{table}
  \caption{Records in Sample.}
  \label{tab:sample}
  \begin{tabular}{lllll}
    \toprule
    Title&Reference Code&LD&Group of Fonds&Dates\\
    \midrule
    Juízo da Índia e Mina&\href{https://digitarq.arquivos.pt/details?id=4208377}{PT/TT/JIM}&Fonds&CA&1700-1833\\
    Autos cíveis de petição [...] &\href{https://digitarq.arquivos.pt/details?id=4208380}{PT/TT/JIM/A/0001/00001}&File&CA&1806-1806\\
    Juízo das Justificações [...] &\href{https://digitarq.arquivos.pt/details?id=4211646}{PT/TT/JIM-JJU}& Subfonds& CA&1700-1833\\
    Cartório do escrivão Lino [...] &\href{https://digitarq.arquivos.pt/details?id=4211635}{PT/TT/JIM/E}&Section&CA&1809-1811\\
    Oriente&\href{https://digitarq.arquivos.pt/details?id=4216803}{PT/TT/JIM-JJU/004}&Serie&CA&1749-1833\\
    Maço 5&\href{https://digitarq.arquivos.pt/details?id=4216708}{PT/TT/JIM-JJU/003/0005}&IU&CA&1765-1828\\
    Autos de sentença de [...] & \href{https://digitarq.arquivos.pt/details?id=4216716}{PT/TT/JIM-JJU/003/0005/00008} &File& CA&1816-1817\\
    Extrato do jornal a [...] &\href{https://digitarq.arquivos.pt/details?id=6035139}{PT/TT/JIM/JJE/0001/00038}&Item&CA& 1813-07-12\\
    Procuração de Vasco [...] & \href{https://digitarq.arquivos.pt/details?id=6036136}{PT/TT/JIM/JJE/0002/00035} &Item&CA &1803-01-21\\
    Acção cível de fretes [...] & \href{https://digitarq.arquivos.pt/details?id=4208504}{PT/TT/JIM/A/0008/00012}&File&CA&1803\\
    "Venda de louça preta […]"&\href{https://digitarq.arquivos.pt/details?id=4600649}{PT/TT/EPJS/SF/008/10253}&Item& Companies&1917\\
    Mosteiro de São Bernardo [...] & \href{https://digitarq.adptg.arquivos.pt/details?id=1001112}{PT/ADPTG/MON/MSBP}&Fonds&Monastics&1443-1882\\
    Gabinete da Área de Sines& \href{https://digitarq.adstb.arquivos.pt/details?id=1199334}{PT/ADSTB/AC/GAS}&Fonds&CA&1941-2006\\
    Carta de Antão Santos [...] & \href{https://digitarq.adctb.arquivos.pt/details?id=1057000}{PT/ADCTB/PSS/APASC/00001}&Item&Personal&1937-09-23\\
    Registo de passaportes deferidos & \href{https://digitarq.adbgc.arquivos.pt/details?id=1228231
    }{PT/ADBGC/AC/GCBGC/PAS/077/01140}&IU&DCA&1844-1858\\
    "Um aquário de 2,40x1,00 m" & \href{https://digitarq.arquivos.pt/details?id=4672523}{PT/TT/AS/A/001/000142}&Item&Personal&1913-04-18\\
    Testamento&\href{https://digitarq.adavr.arquivos.pt/details?id=1257880}{PT/ADAVR/NOT/CNVFR2/002/0001/000007}&Item&Notary&1903-08-26\\
    Planta de parte do 3.º [...] &\href{https://digitarq.arquivos.pt/details?id=4684278}{PT/PNA/DGFP/0002/0001/00047/00002}&Item&CA&1914\\
    Carta náutica, de João [...] &\href{https://digitarq.arquivos.pt/details?id=4162626}{PT/TT/CRT/198}&Item&Collections&1620-1640\\
    Bula "Romani Pontificis" [...] &\href{https://digitarq.arquivos.pt/details?id=6843190}{PT/TT/BUL/0003/06}&Item&Collections&1523-03-12\\
    Processo de Jerónima da Cruz & \href{https://digitarq.arquivos.pt/details?id=2362242}{PT/TT/TSO-IE/021/00203}&File&CA&1547-03-12\\
    Desembargo do Paço&\href{https://digitarq.arquivos.pt/details?id=4167317}{PT/TT/DP}&Fonds&CA&1610-1833\\
    Inventário por óbito de [...]  &\href{https://pesquisa.adporto.arquivos.pt/details?id=1014990}{PT/ADPRT/JUD/TCVNG/077/00053}&File&Judicial&1973-1974\\
    Registo de batismo&\href{https://digitarq.adbja.arquivos.pt/details?id=1040274}{PT/ADBJA/PRQ/AJT01/001/0013}&IU&Parish&1862\\
    Notas para escrituras diversas &\href{https://digitarq.adstr.arquivos.pt/details?id=1013873}{PT/ADSTR/NOT/07CNABT/001/0023}&IU&Notary&1703\\
  \bottomrule
  \multicolumn{5}{c}{\small LD, IU, CA, and DCA stand for Level of Description, Installation Unit, Central Administration, and Decentralized Central}\\
  \multicolumn{5}{c}{\small Administration, respectively.}
\end{tabular}
\end{table}

\subsection{Representation of ``Juízo da Índia e Mina''}
\label{sec:Representation}

\begin{figure}[ht]
\centering
    \includegraphics[scale=0.36]{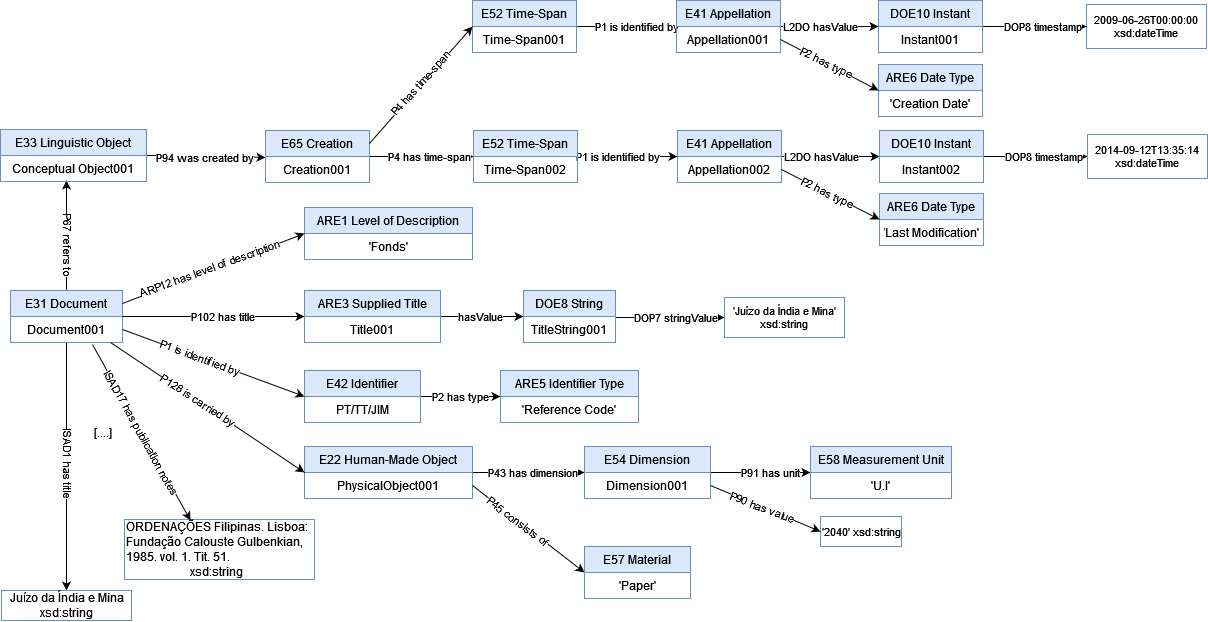}
\caption{Representation example of ``Juízo da Índia e Mina''.}
\label{fig:mappingExample}
\end{figure}

We focus here on the representation of the first record in the sample, ``Juízo da Índia e Mina'', with PT/TT/JIM as reference code -- Figure~\ref{fig:mappingExample}. This is a Unit of Description at the level of the fonds, and has information that pertains to all documents in the fonds. This record has, therefore, a description with substantial detail. 
The ISAD(G) elements that constitute the record have information that translates directly to ArchOnto individuals and properties, and other information that needs to be atomized to become part of the ArchOnto representation. We started with the ISAD(G) elements that have a direct representation in the new model.

The record we are representing is the description of a document, and the document as a whole is present as an object of the \textit{E31 Document} class. Note that ``Document'' will be used for Units of Description (UD) at different levels in the archives hierarchy, namely the fonds level in this example. In ISAD(G), UD is an abstract concept, used for units at different levels and ``Document'' is a specific level, while in CIDOC CRM parlance all UD correspond to documents with different levels of aggregation.

The first ISAD(G) element to be considered is the \textit{Level of Description}, represented in ArchOnto through the \textit{ARE1 Level of Description} class. This class is a subclass of \textit{E55 Type} and is associated with a document (\textit{E31 Document}) through the property \textit{ARP12 has level of description}. 

The document title is the next core element represented through ArchOnto. In the example, the document title is a supplied title, and an object of class \textit{ARE3 Supplied Title} is used to represent the title. This class, a subclass of \textit{E35 Title}, indicates the type of title used in the document record. In addition to this type of title, assigned by an archivist, in ArchOnto, it is possible to represent a formal title, originated in the document --  \textit{ARE2 Formal Title}. The representation of the title also uses the \textit{DOE8 String} class of the DataObject ontology to define and validate the actual string of characters in the title.

One of the main identifiers of a document in ISAD(G), the reference code, is also represented in ArchOnto, through the class \textit{E42 Identifier}, which links to the document through the property \textit{P1 is identified by}. ISAD(G) allows several kinds of identifiers. The fact that this is a reference code is visible in the association to an object of the \textit{ARE5 Identifier Type} class. \textit{ARE5} is a subclass of \textit{E55 Type} and is linked to \textit{E42 Identifier} through the property \textit{P2 has type}.

In Figure~\ref{fig:mappingExample}, it is also possible to observe the representation of the dimension and support of the document, which originated in the representation of the physical documents owned by the ANTT. Archival objects include several kinds of documents, in digital or physical format. The latter may or may not have associated digital representations. %RESPOSTA REVISOR 
As these ISAD(G) elements concern the physical part of the document, they are linked to the document via an instance of the \textit{E22 Human-Made Object} class. The document support is represented with class \textit{E57 Material} and is linked to the physical object through the property \textit{P45 consists of}. On the other hand, the dimension is represented by \textit{E54 Dimension} and has two associated values, the dimension measurement unit --- \textit{E58 Measurement Unit} --- and its value. These two values are linked to the dimension through  properties \textit{P91 has unit} and \textit{P90 has value}, respectively.

In addition to the physical document properties, there is ISAD(G) information regarding the corresponding conceptual object, which is represented in Figure~\ref{fig:mappingExample} through the \textit{E33 Linguistic Object} class. In this figure, we have the representation of the creation date of the description of the archival document. Several classes are necessary to validate the data. The date representation starts with the presentation of the creation event of the archival record (\textit{E65 Creation}), followed by two different \textit{E52 Time-Span}, one for the creation of the record and the other to express the last modification. The time-span has an appellation (\textit{E41 Appellation}) that is composed of an \textit{DOE10 Instant}. In the Instant, with a timestamp, the date is typed and validated through the DataObject. The dates have a format complying with  the dateTime datatype --- xsd: dateTime, with pattern “YYYY-MM-DDThh:mm:ss”.

The complete information in the ISAD(G) elements for this fonds is captured, in this representation, with the ISAD Ontology. This is a simple, auxiliary ontology, where the textual elements are kept whole for migrated records.  For simplification purposes, in Figure~\ref{fig:mappingExample}, we only show the representation of the title and publication notes.

Information from more verbose ISAD(G) elements will also be atomized in ArchOnto. To this end, several ArchOnto classes were identified, capturing concepts that are present in the ISAD(G) elements, including events (\textit{E5 Event}), people (\textit{E21 Person}), places (\textit{E53 Place}) and kinds of ships (\textit{E55 Type}).

%%%%%%%%%%%%
\subsection{Properties and classes used in the manual representation}
\label{sec:Analysis}
With the representation of selected records, it was possible to observe the relative use of the classes and properties in ArchOnto. Table~\ref{tab:mostUsedClasses} and Table~\ref{tab:mostUsedProperties} provide the number of times each class and property was used in the manual representation of the 25-record sample.

\begin{table} [ht]
    \caption{Occurrences of classes in the migration of the sample records}
    \label{tab:mostUsedClasses}
       \begin{minipage}[t]{.5\linewidth}
       \vspace{0pt}
        \centering
        \begin{tabular}{lll}
        \toprule
        Ontology&Classes&Nr\\
        \midrule
        CIDOC CRM&E41 Appellation&247\\
        DataObject&DOE8 String&209\\
        CIDOC CRM&E52 Time-Span&106\\
        CIDOC CRM&E42 Identifier&70\\
        CIDOC CRM&E31 Document&69\\
        CIDOC CRM&E21 Person&63\\
        CIDOC CRM&E22 Man-Made Object&61\\
        DataObject&DOE17 PersonName&53\\
        DataObject&DOE11 Interval&52\\
        DataObject&DOE10 Instant&49\\
        CIDOC CRM&E53 Place&45\\
        N-Ary&PC14 Carried Out By&45\\
        ArchOnto&ARE3 Supplied Title&38\\
        CIDOC CRM&E35 Title&33\\
        CIDOC CRM&E54 Dimension&32\\
        ArchOnto&ARE8 Role Type&30\\
        CIDOC CRM&E74 Group&30\\
        ArchOnto&ARE11 Documentary Typology&25\\
        CIDOC CRM&E12 Production&25\\
        CIDOC CRM&E33 Linguistic Object&25\\
        CIDOC CRM&E65 Creation&25\\
        CIDOC CRM&E7 Activity&21\\
        CIDOC CRM&E55 Type&19\\
        CIDOC CRM&E32 Authority Document&13\\
        CIDOC CRM&E57 Material&13\\
        
      \bottomrule
    \end{tabular}
    \end{minipage}%
    \begin{minipage}[t]{.5\linewidth}
    \vspace{0pt}
      \centering
        \begin{tabular}{lll}
        \toprule
        Ontology&Classes&Nr\\
        \midrule
        CIDOC CRM&E58 Measurement Unit&12\\
        CIDOC CRM&E67 Birth&9\\
        ArchOnto&ARE1 Level of Description&8\\
        ArchOnto&ARE2 Formal Title&8\\
        CIDOC CRM&E3 Condition State&8\\
        CIDOC CRM&E10 Transfer of Custody&8\\
        CIDOC CRM&E56 Language&8\\
        ArchOnto&ARE6 Date Type&5\\
        CIDOC CRM&E66 Formation&5\\
        ArchOnto&ARE5 Identifier Type&4\\
        CIDOC CRM&E5 Event&4\\
        ArchOnto&ARE4 Extension&3\\
        ArchOnto&ARE14 Place Type&3\\
        ArchOnto&ARE15 Acquisition Type&3\\
        CIDOC CRM&E68 Dissolution&3\\
        ArchOnto&ARE9 Date Certainty&2\\
        ArchOnto&ARE16 Event Type&2\\
        CIDOC CRM&E69 Death&2\\
        ArchOnto&ARE13 Subject Type&1\\
        DataObject&DOE9 Approximate&1\\
        CIDOC CRM&E9 Move&1\\
        CIDOC CRM&E36 Visual Item&1\\
        CIDOC CRM&E85 Joining&1\\
        CIDOC CRM&E86 Leaving&1\\
        CIDOC CRM&E96 Purchase&1\\
      \bottomrule
        \end{tabular}
    \end{minipage}
\end{table}

\begin{table} [ht]
    \caption{Occurrences of properties in the migration of the sample records}
    \label{tab:mostUsedProperties}
       \begin{minipage}[t]{.5\linewidth}
       \vspace{0pt}
        \centering
        \begin{tabular}{lll}
        \toprule
        Ontology&Property&Nr\\
        \midrule
        DataObject&DOP7 string value&418\\
        CIDOC CRM&P1 is identified by&313\\
        CIDOC CRM&P2 has type&223\\
        DataObject&DOP5 name&106\\
        DataObject&DOP2 end date value&102\\
        DataObject&DOP6 start date value&102\\
        CIDOC CRM&P4 has time-span&101\\
        DataObject&DOP8 timestamp&98\\
        CIDOC CRM&P70 documents&86\\
        CIDOC CRM&P102 has title&79\\
        CIDOC CRM&P3 has note&74\\
        CIDOC CRM&P90 has value&70\\
        CIDOC CRM&P14.1 in the role of&49\\
        ArchOnto&ARP12 has level of description&48\\
        CIDOC CRM&P67 refers to&47\\
        N-Ary&P01 has domain&47\\
        ISAD Ontology&ISAD9 has scope&36\\
        CIDOC CRM&P43 has dimension&35\\
        CIDOC CRM&P91 has unit&34\\
        CIDOC CRM&P128 is carried by&33\\
        CIDOC CRM&P45 consists of&32\\
        CIDOC CRM&P72 has language&30\\
        CIDOC CRM&P106 is composed of&27\\
        CIDOC CRM&P94 was created by&25\\
        CIDOC CRM&P108 was produced by&25\\
        N-Ary&P02 has range&23\\
        CIDOC CRM&P107 has current or former [...]&20\\
        ISAD Ontology&ISAD10 has conditions [...]&18\\
        ArchOnto&ARP8 upper level&16\\
        ArchOnto&ARP9 lower level&16\\
        ISAD Ontology&ISAD18 has note&16\\
        CIDOC CRM&P130 features are also found on&15\\
        CIDOC CRM&P46 is composed of&14\\
        ISAD Ontology&ISAD15 has related unit of [...]&14\\
        ISAD Ontology&ISAD7 has administrative [...]&12\\
        ISAD Ontology&ISAD19 has system of [...]&12\\
        ISAD Ontology&ISAD24 has conditions [...]&12\\
        CIDOC CRM&P11 had participant&11\\
        CIDOC CRM&P89 falls within&11\\
        CIDOC CRM&P129 is about&11\\
        CIDOC CRM&P7 took place at&10\\
        CIDOC CRM&P98 brought into life&10\\
      \bottomrule
    \end{tabular}
    \end{minipage}%
    \begin{minipage}[t]{.5\linewidth}
    \vspace{0pt}
      \centering
        \begin{tabular}{lll}
        \toprule
        Ontology&Property&Nr\\
        \midrule
        ISAD Ontology&ISAD8 has archival history&10\\
        ISAD Ontology&ISAD25 has conditions governing use&10\\
        ISAD Ontology&ISAD26 has immediate source of [...]&10\\
        CIDOC CRM&P14 carried out by&8\\
        CIDOC CRM&P28 custody surrendered by&8\\
        CIDOC CRM&P29 custody received by&8\\
        CIDOC CRM&P30 transfered custody of&8\\
        CIDOC CRM&P97 from father&8\\
        ISAD Ontology&ISAD16 has existence and location [...]&8\\
        ISAD Ontology&ISAD17 has publication note&8\\
        CIDOC CRM&P74 has current or former residence&6\\
        CIDOC CRM&P96 by mother&6\\
        ISAD Ontology&ISAD20 has physical characteristics [...]&6\\
        CIDOC CRM&P5 consists of&5\\
        CIDOC CRM&P12 was present at&5\\
        CIDOC CRM&P95 was formed by&5\\
        CIDOC CRM&P20 had specific purpose&4\\
        CIDOC CRM&P50 has current keeper&4\\
        CIDOC CRM&P53 has former or current location&4\\
        CIDOC CRM&P134 continued&4\\
        ISAD Ontology&ISAD27 has accruals&4\\
        CIDOC CRM&P44 has condition&3\\
        CIDOC CRM&P99 was dissolved by&3\\
        CIDOC CRM&P122 borders with&3\\
        DataObject&DOP1 approximate date value&2\\
        CIDOC CRM&P17 was motivated by&2\\
        CIDOC CRM&P48 has preferred identifier&2\\
        CIDOC CRM&P100 was death of&2\\
        CIDOC CRM&P121 overlaps with&2\\
        CIDOC CRM&P24 changed ownership through&1\\
        CIDOC CRM&P25 moved by&1\\
        CIDOC CRM&P26 moved to&1\\
        CIDOC CRM&P49 has former or current keeper&1\\
        CIDOC CRM&P54 has current peranent location&1\\
        CIDOC CRM&P71 is listed in&1\\
        CIDOC CRM&P143 joined&1\\
        CIDOC CRM&P144 joined with&1\\
        CIDOC CRM&P145 separated&1\\
        CIDOC CRM&P146 separated from&1\\
        CIDOC CRM&P151 was formed from&1\\
        CIDOC CRM&P173 ends with or after the start of&1\\
        CIDOC CRM&P183 starts after the end of&1\\
      \bottomrule
        \end{tabular}
    \end{minipage}
\end{table}

As expected, the CIDOC CRM classes and properties are prevalent in Table~\ref{tab:mostUsedClasses} and Table~\ref{tab:mostUsedProperties}, indicating that CIDOC CRM is ArchOnto's base ontology.

Among the classes used, some stand out, being used more intensively, noticeably the \textit{E41 Appellation}, \textit{DOE8 String}, and \textit{E52 Time-Span}. \textit{E41 Appellation} is a unifying object for the names of different entities, which justifies its frequent use. The \textit{DOE8 String} class is also widely used because it is used to insert and verify textual elements, such as titles, identifiers, and names (such as the names of places, activities, groups). In the extensions of CIDOC CRM for archives, the classes for Titles (\textit{ARE3 Supplied Title}), Role Type (\textit{ARE8 Role Type}), and Documentary Typology (\textit{ARE11 Documentary Typology}) also appear repeatedly.

Regarding properties, we have high numbers for properties such as \textit{DOP7 string value}, used to express the string values, \textit{P1 is identified by}, associating objects with multiple identifiers, and \textit{P2 has type}, used for associating individuals with their types. In the extensions to CIDOC CRM, \textit{ARP12 has level of description} is used for all 25 records as expected, and to represent the level of description of documents related to the 25 records in the sample. Properties from the ISAD Ontology are used to represent information that comes from textual fields. Properties from the Data Object ontology are also extensively used, dealing with literal values. 

From 2652 properties used to represent the 25 records, 1498 are from CIDOC CRM, 828 from DataObject, 176 from ISAD Ontology, 80 from ArchOnto, and 70 from N-ary. With this, we can observe the importance and multidisciplinarity that CIDOC CRM presents and how the ontologies created to represent the archival documents complement it.

%%%%%%%%%%%%%%%%%%%%%%%%%%%%%%%%%%%%%%%%%%%%%%%%%%%%%%%
\section{Rules for the Automatic Migration of ISAD(G) records to ArchOnto}
\label{sec:migration}

Based on the experience of manually representing the 25-records sample, and envisioning the automatic migration of the existing records to ArchOnto, we defined rules to map each ISAD(G) element. These rules are presented in Table~\ref{tab:isadArchOnto} using the Mapping Description Language~\cite{mapping}. %In this table where is used ``\{\}'' is intended to be assigned identifiers and where \$ is intended to refer to the element that same id. 

\begin{table}
\small
  \caption{From ISAD(G) to ArchOnto: migration rules}
  \label{tab:isadArchOnto}
  \begin{tabular}{cll}
    \toprule
    Rule No. & ISAD(G) & ArchOnto Path\\
    \midrule
    1&ISAD\{D1\} & \texttt{E31 Document\{=D1\}};\\
    && \texttt{\$D1 -> P128 is carried by -> E22 Human-Made Object\{=HMO1\}};\\
    && \texttt{\$D1-> P67 refers to -> E33 Linguistic Object\{=LO1\}};\\
    \midrule
    2&\$D1->Description Level\{DL\} &\texttt{\$D1 -> ARP12 has level of description -> ARE1 Level of }\\
    &&\texttt{Description\{=DL\}};\\
    \midrule
    3&\$D1->Reference Code\{RC\} & \texttt{\$D1 -> P1 is identified by -> E42 Identifier \{=RC\} ->} \\
    && \texttt{P2 has type -> ARE5 Identifier Type \{='Reference Code'\}};\\
    \midrule
    4&\$D1->Title\{T\} & \texttt{\$D1 -> P102 has title -> E35 Title -> DOP7 stringValue -> T};\\
    \midrule
    5&\$D1->Formal Title\{FT\} & \texttt{\$D1 -> P102 has title -> ARE2 Formal Title -> DOP7 stringValue}\\
    &&\texttt{-> FT};\\
    \midrule
    6&\$D1->Supplied Title\{ST\} & \texttt{\$D1 -> P102 has title -> ARE3 Supplied Title -> DOP7 stringValue}\\
    &&\texttt{-> ST};\\
    \midrule
    7&\$D1->Production Date\{SD, ED\} & \texttt{\$HMO1 -> P108 was produced by -> E12 Production ->}\\
    &&\texttt{P4 has time-span -> E52 Time-Span -> P1 is identified by -> }\\
    &&\texttt{E41 Appellation -> L2DO hasValue -> DOE11 Interval \{=INT1\}};\\ 
    &&\texttt{\$INT1 -> DOP6 startDateValue -> SD};\\
    && \texttt{\$INT1 -> DOP2 endDateValue -> ED};\\ 
    \midrule
    8&\$D1->Production Date\{PD\} & \texttt{\$HMO1 -> P108 was produced by -> E12 Production ->}\\
    &&\texttt{P4 has time-span -> E52 Time-Span -> P1 is identified by ->}\\
    && \texttt{E41 Appellation -> L2DO hasValue -> DOE10 Instant ->}\\
    &&\texttt{DOP8 timestamp -> PD};\\ 
    \midrule
    9&\$D1->Dimension\{DIM\} & \texttt{\$HMO1 -> P43 has dimension -> E54 Dimension \{=DIM1\}};\\
    && \texttt{\$DIM1 -> P91 has unit -> E58 Measurement Unit};\\
    && \texttt{\$DIM1 -> P90 has value -> DIM};\\
    \midrule
    10&\$D1->Extension\{EXT\} &  \texttt{\$HMO1 -> P43 has dimension -> ARE4 Extension\{=E1\}};\\
    && \texttt{\$E1 -> P91 has unit -> E58 Measurement Unit};\\
    && \texttt{\$E1 -> P90 has value -> EXT};\\
    \midrule
    11&\$D1->Support\{SP\}&\texttt{\$HMO1 -> P45 consists of -> E57 Material\{=SP\}};\\
    \midrule
    12&\$D1->Language\{LG\}& \texttt{\$LO1 -> P72 has language -> E56 Language\{=LG\}};\\
    \midrule
    13&\$D1->Physical Location\{PL\}& \texttt{\$D1 -> P1 is identified by -> E42 Identifier \{=PL\} ->}\\
    &&\texttt{P2 has type -> ARE5 Identifier Type \{='Physical Location'\}};\\
    \midrule
     14&\$D1->Original Numbering\{ON\}& \texttt{\$D1 -> P1 is identified by -> E42 Identifier \{=ON\} ->}\\
    &&\texttt{P2 has type -> ARE5 Identifier Type \{='Original Numbering'\}};\\
    \midrule
     15&\$D1->Previous Location\{PreL\}& \texttt{\$D1 -> P1 is identified by -> E42 Identifier \{=PreL\} ->}\\
    &&\texttt{P2 has type -> ARE5 Identifier Type \{='Previous Location'\}};\\
    \midrule
    16&\$D1->Creation Date\{CD\}& \texttt{\$LO1 -> P94 was created by -> E65 Creation -> P4 has time-span}\\
    &&\texttt{-> E52 Time-Span -> P1 is identified by -> E41 Appellation ->} \\
    &&\texttt{L2DO hasValue -> DOE10 Instant  \{=INST1\}};\\
    &&\texttt{\$INST1 -> DOP8 timestamp -> CD};\\
    &&\texttt{\$INST1 -> P2 has type -> ARE6 Date Type \{='Creation Date'\}};\\
    %&&\texttt{\$INT2 -> DOP6 startDateValue -> SCD};\\
   % &&\texttt{\$INT2 -> DOP2 endDateValue -> ECD};\\
    \midrule
    17&\$D1->Parent Record\{PR\}&\texttt{\$D1 -> P165 incorporates -> \$PR};\\
  \bottomrule
  \multicolumn{3}{c}{\small Conjugation of rules is done with ´;´. V outputs the value of the variable V; \$V is the element identified by the value of the variable V;}\\
  \multicolumn{3}{c}{\small and \{=V\} is the assignment of the value of V. In the assignment, if V is not yet defined, it takes the value of `V'. The assignment of a }\\
  \multicolumn{3}{c}{\small  literal is done with {=`literal value'}.}\\
\end{tabular}
\end{table}

The migration rules are composed of two parts. The first part represents the ISAD(G) representation, and the second represents the ArchOnto Path. To read the rules, it is essential to consider that, in the ArchOnto Path, classes and properties of the ontology are used, as well as symbols that represent what was already present in ISAD(G) -- represented by ``\{\}'' in the first part of the rule, as can be seen, for example, in \{=D1\}.

To create a document, we can use the first rule, which is applied for each ISAD(G) record associated with a physical part and a conceptual part. The physical part describes aspects of the physical object, such as dimension and support. It is important to emphasize that, in digital objects, the physical object is not palpable. The conceptual part, on the other hand, includes aspects such as the language in which the document is written. 
The path declares that a document (\textit{E31 Document}) is carried by (\textit{P128 is carried by}) a Human-Made Object (\textit{E22 Human-Made Object}), and the same document refers to (\textit{P67 refers to}) a Linguistic Object (\textit{E33 Linguistic Object}). 

To represent the title of a document, we can use rules 4, 5, and 6, depending on the type of title available on the document. When a document has a formal title type, Rule~5 must be taken into account. If the document has a title with the type supplied, Rule~6 must be used. If the type of title is not present in the document, Rule~4 should be used. These three rules follow similar patterns since each title is always associated with a document (\textit{E31 Document}), through the property \textit{P102 has title}, regardless of its type.
The following path shows how to represent document with a Formal Title (\textit{ARE2 Formal Title}). A document (\$D1) has a formal title (\textit{P102 has title -> ARE2 Formal Title}) which is represented by a string (\textit{DOP7 stringValue}). 

To represent the date of creation of a record, rule 16, we need to define two different instants. The first one corresponds to the creation of the description, and the second to the more recent modification.
The path describes a Linguistic Object (\$LO1) created by the event of creation (\textit{P94 was created by -> E65 Creation}), which has two time-spans (\textit{P4 has time-span -> E52 Time-Span}) identified by an appellation with an instant (\textit{P1 is identified by -> E41 Appellation -> L2DO hasValue -> DOE11 Instant}). This instant has a timestamp (\textit{DOP8 timestamp}) and a date type, depending on if it is the Creation Date or the Last Modification of the record (\textit{P2 has type -> ARE6 Date Type \{='Creation Date'\}} or \textit{\{='Last Modification'\}}).

In addition to the rules presented in the table, there are also rules for the use of the ISAD Ontology. All properties of the ISAD Ontology are used in the same way, and all properties have the \textit{E31 Document} class as domain and the xsd:string as range. For example, \textit{E31 Document -> ISAD1 has title -> xsd:string}.

%%%%%%%%%%%%%%%%%%%%%%%%%%%%%%%%%%%%%%%%%%%%%%%%%%%%%%%
\section{Discussion of Results}
\label{sec:Discussion}
The CIDOC CRM is a model designed for museums and it explored to represent the main concepts for other cultural heritage domains.   
This work is centered on the development of ArchOnto, an ontology to support a new archival information system at the Portuguese National Archives. The ontology is based on the CIDOC CRM for the core concepts, includes extensions to deal with archival-specific elements, and incorporates some additional ontologies to ease the transition from legacy records and deal with practical aspects of the validation of literal values. The CIDOC CRM can take care of the representation for most of the information in the archival model but was not suitable to represent all the major elements in the ISAD(G) standard. This led to the creation of additional classes and  properties, some of them organised as sub-ontologies to complete the model.

Although the model is designed to support a new workflow for archival description, the existing records constitute a huge resource with valuable information generated by specialists. This work explores the representation of existing records, from the legacy DigitArq database, using the ArchOnto ontology. It is visible that CIDOC CRM contributes with most of the core concepts, but also that extensions are required to deal with archival records.

Working with information from existing records it was possible to conclude, for instance, that the \textit{E55 Type} class is extensively used, due to the wide range of concepts it supports. The examples made it clear that a flat use of \textit{E55 Type} would provide only very general information on the category of object that it would classify, limiting the use of ``type'' as an effective means for enforcing controlled vocabularies. 
One of the extensions in ArchOnto was therefore a set of subclasses of \textit{E55 Type} to specialize types that are common in archives, such as those for level of description, dates, or document types, most of them corresponding to controlled vocabularies in the domain of archives. So far, a list of 15 subclasses has been considered relevant by the archive specialists.

%falta completar esta tabela
\begin{table}[]
  \caption{Status of the migration of ISAD elements}
  \label{tab:migrationStatus}
  \begin{tabular}{lc}
    \toprule
    Element & Representation in ArchOnto \\
    \midrule
    \textbf{Identity Statement Area} &\\
    1.1. Reference code (mandatory) & C \\
    1.2. Title (mandatory) & C \\
    1.3. Dates (mandatory) & C\\
    1.4. Level of description (mandatory)& C \\
    1.5. Extent and medium of the UoD (mandatory)& C \\
    \textbf{Context Area} &\\
    2.1. Name of Creator (mandatory) &C\\
    2.2. Administrative/biographical history & PC \\
    2.3. Archival history & PC \\
    2.4. Immediate source of acquisition or transfer & PC \\
    \textbf{Content and Structure Area} &\\
    3.1. Scope and content& PC \\
    3.2. Appraisal, destruction and scheduling information& PC\\
    3.3. Accruals& NS\\
    3.4. System of Arrangement& NS\\
    \textbf{Conditions of Access and Use Area}& \\	
    4.1. Conditions governing access&NS\\
    4.2. Conditions governing reproduction& NS\\
    4.3. Language/scripts of material& C \\
    4.4. Physical characteristics and technical requirements& C\\
    4.5. Finding aids &NS\\
    \textbf{Allied Materials Area}  &\\
    5.1. Existence and location of originals &C\\
    5.2. Existence and location of copies &C\\
    5.3. Related units of description &PC\\
    5.4. Publication note &PC\\
    \textbf{Notes Area} &\\
    6.1. Note &PC\\
    \textbf{Description control area} &\\	
    7.1. Archivist’s Note &PC\\
    7.2. Rules or conventions &NS\\
    7.3. Date of description &C\\
  \bottomrule
  \multicolumn{2}{c}{\small C, PC, and NS stand for Complete, Partially Complete, and Not Started, respectively.}
\end{tabular}
\end{table}

On the other hand, the examples have led us to observe that the most descriptive elements, such as ``Scope and content'' and ``Administrative/ Biographical history'', include information on e.g. events, people, organizations, entities, that the archive specialists consider important enough to be captured in a structured manner, such as allowed by the CIDOC CRM. 
Therefore, the representation of these elements in ArchOnto is not processed as a direct mapping. Still, the identification of relevant entities that are common in the textual information of these elements has already started. Relevant concepts are those of person and group (that can be represented with classes \textit{E21 Person} and \textit{E74 Group}, respectively), event (\textit{E5 Event}), date (\textit{E54 Time-Span}), and place (\textit{E54 Place}). These are clearly central concepts in CIDOC CRM as well. For the migration of the textual elements that can provide such entities, we will use natural language processing tools to analyze the content and extract the concepts. The inference of ontological relations will also be considered.

%\subsection{Manual representation and migration status}
%\label{sec:Status}
It is interesting to assess the extent to which the ISAD(G) elements in existing records have been mapped to ArchOnto. Table \ref{tab:migrationStatus} list the elements, grouped by their area in ISAD(G) and shows the status of their representation in the manual representation using ArchOnto. ``C'' stands for ``Complete'', ``PC'' for ``Partially Complete'' and ``NS'' for ``Not Started''. For elements marked as not started, although they are not yet directly represented, the CIDOC CRM concepts to represent the information present in the ISAD(G) description are already identified.

When analyzing existing records in DigitArq, it was possible to observe many null or blank values in ISAD(G) elements. Further analysis led to the conclusion that this is a result of two of the principles adopted in ISAD(G): description proceeds from general to more specific, and there should be no repetition of information~\cite{InternationalCouncilonArchives2000}. An element filled at fonds level, for instance, is not repeated for a series that is part of the fonds -- but it may be overridden by more specific information. As the standard has a hierarchical structure, it is only mandatory to fill some of the elements at the top level. With this, a selected record inherits the description corresponding to the level immediately above, in case a new value is not provided in its own record. 
As the systematic migration becomes complete, inferences will be performed by the applications that explore the knowledge graph. For missing values in ISAD(G) elements, the corresponding records up the hierarchy should be considered.

%%%%%%%%%%%%%%%%%%%%%%%%%%%%%%%%%%%%%%%%%%%%%%%%%%%%%%%
\section{Conclusions}
\label{sec:Conclusions}

This work is part of a broader endeavour that started with a new vision for the Portuguese Archives, one where the information in the archives, which is increasingly requested by citizens for a diversity of purposes, is explored in its connections to the wider information world. This plan requires a vast investment in conceptualization, modelling, prototyping, user testing, development and evaluation.

This work reports on the activities related to the definition of a linked open data model for archives. It started with a study of the CIDOC CRM standard by the archival specialists. This provided insight into the level of detail at which current models were exploring the representation of cultural assets. On the other hand, the computer science team had previous work based on ISAD(G) and were familiar with the Digitarq system used by the Portuguese archives. Their path included a thorough study of CIDOC CRM and a systematic analysis of the contents of existing records.

This starting point allowed the continuation of work in three parallel lines: the definition of the data model, the selection of a technology stack to support the knowledge graph, and the definition of user profiles and interfaces for user access and interaction. These tasks are reaching their conclusion, giving way to those that rely on their results. 

Building on the data model, an automatic migration process is ongoing. Rules are already in place for most of the ISAD(G) elements for which the mapping into ArchOnto is straightforward. Extra effort is required for elements that are rich in textual content an will require more elaborate natural language processing for extraction.

Building on the selected technology stack, a triple store is being built to incorporate the results of the migration and to support the prototype applications that include generic search interfaces and environments for archival activities such as collection organisation, description and conservation. 

Moreover, the overall plan also includes two additional aspects: text extraction from digital documents and knowledge propagation. The former accounts for the exploration of the textual content of the born-digital documents. The goal is to ease their description by generating automated elements of description to be validated and completed by the archivists. The latter proceeds to connect the information in the archives to other sources, providing archival users with a glimpse of the relations between entities in the archives and their counterparts in more informal data sources.

One aspect that is very relevant in this plan is the conformity to the standards and recommendations by the archival professional bodies, namely the International Council on Archives (ICA). ICA is developing a linked data model (RiC-CM). By the time our project started, in January 2019, little information was available about this new model. At that time, we started to explore CIDOC CRM, a model originated in museums, with a long period of development that makes it a stable model. The study of RiC-CM and its connections to CIDOC CRM are in the scope of future work related to this project. 
After all, an overarching goal for cultural heritage platforms is to link objects from different sources. Regardless of what actual model you build a specific system on, they are expected to interoperate and be equipped with pathways for communication that promote seamless perusal of cultural heritage assets across institutions and technologies.

\begin{acks}
This work is financed by National Funds through the Portuguese funding agency, FCT -- Fundação para a Ciência e a Tecnologia within project DSAIPA/DS/0023/2018. Inês Koch is also financed by National Funds through the Portuguese funding agency, FCT – Fundação para a Ciência e a Tecnologia within the research grant 2020.08755.BD.
\end{acks}

\bibliographystyle{ACM-Reference-Format}
\bibliography{episa_JOCCH}

\newpage
\appendix

\section{THE CLASS HIERARCHY WITHIN ARCHONTO’S ONTOLOGIES}

\begin{table}[ht]
  \caption{The class hierarchy within ArchOnto’s ontologies}
  \label{tab:classes_ArchOnto}
  \begin{tabular}{ll}
    \toprule
    Ontology & Classes\\
    \midrule
    CIDOC CRM  & E35 Title\\
    Archonto& --- ARE2 Formal Title \\
    Archonto& --- ARE3 Supplied Title\\
    CIDOC CRM & E39 Actor\\
    Archonto & --- ARE12 Organisation \\
    CIDOC CRM & E54 Dimension\\
    Archonto& --- ARE4 Extension \\
    CIDOC CRM & E55 Type\\
    Archonto & --- ARE1 Level of Description\\
    Archonto& --- ARE5 Identifier Type \\
    Archonto& --- ARE6 Date Type \\
    Archonto& --- ARE7 Name Type \\
    Archonto& --- ARE8 Role Type \\
    Archonto& --- ARE9 Date Certainty \\
    Archonto& --- ARE11 Documentary Typology \\
    Archonto& --- ARE13 Subject Type \\
    Archonto& --- ARE14 Place Type\\
    Archonto& --- ARE15 Acquisition Type \\
    Archonto& --- ARE16 Event Type \\
    DataObject& DOE1 DataObject  \\
    DataObject& --- DOE2 AuthorityFile    \\
    DataObject& --- DOE3 Boolean  \\
    DataObject& --- DOE4 Date  \\
    DataObject& --- --- DOE9 Approximate \\
    DataObject& --- --- DOE10 Instant \\
    DataObject& --- --- DOE11 Interval \\
    DataObject& --- DOE5 Decimal \\
    DataObject& --- DOE6 GeospatialCoordinates \\
    DataObject& --- --- DOE12 Latitude \\
    DataObject& --- --- DOE13 Longitude \\
    DataObject& --- --- DOE14 Polygon  \\
    DataObject& --- DOE7 Integer \\
    DataObject& --- DOE 8String \\
    DataObject& --- --- DOE15 AuthorityString \\
    DataObject& --- --- --- DOE17 PersonName \\
    DataObject& --- --- DOE16 RegexString  \\
    N-ary & PC0 CRM Property  \\
    N-ary & --- PC14 Carried Out By  \\
  \bottomrule
\end{tabular}
\end{table}

\newpage

\section{THE PROPERTY HIERARCHY WITHIN ARCHONTO’S ONTOLOGIES}

\begin{table}[ht]
  \caption{The property hierarchy within ArchOnto’s ontologies}
  \label{tab:archontoProperties}
  \begin{tabular}{llll}
    \toprule
    Ontology & Property & Domain & Range\\
    \midrule
    ArchOnto & ARP8 upper level & ARE1 Level of Description & ARE1 Level of Description\\
    ArchOnto &ARP9 lower level & ARE1 Level of Description & ARE1 Level of Description\\
    Link2DataObject & L2DO hasValue & E1 CRM Property & DOE1 DataObject\\
    N-ary & P01 has domain & PC0 CRM Property & E1 CRM Entity  \\
    N-ary & P02 has range & PC0 CRM Property & E1 CRM Entity  \\
    DataObject & DOP1 approximateDateValue & DOE9 Approximate & xsd:dateTime\\
    DataObject & DOP2 endDateValue & DOE11 Interval & xsd:dateTime\\
    DataObject & DOP3 fileLocation & DOE15 AuthorityFile & xsd:string\\
    DataObject & DOP4 hasRegex & DOE16 RegexString & xsd:string\\
    DataObject & DOP5 name & DOE17 PersonName & xsd:string\\
    DataObject & DOP6 startDateValue & DOE11 Interval & xsd:dateTime\\
    DataObject & DOP7 stringValue & DOE8 String & xsd:string\\
    DataObject & DOP8 timestamp & DOE10 Instant & xsd:dateTime\\
    CIDOC CRM & P2 has type & E1 CRM Entity & E55 Type\\
    ArchOnto&  --- ARP12 has level of description & E31 Document & ARE1 Level of Description\\
    CIDOC CRM & P3 has note & E1 CRM Entity & ---------\\
    ISAD Ontology & --- ISAD1 has title  & E31 Document & xsd:string\\
    ISAD Ontology & --- ISAD2 has level of description & E31 Document & xsd:string\\
    ISAD Ontology& --- ISAD3 has reference code & E31 Document & xsd:string\\
    ISAD Ontology & --- ISAD4 has type of title & E31 Document & xsd:string\\
    ISAD Ontology & --- ISAD5 has date & E31 Document & xsd:string\\
    ISAD Ontology & --- ISAD6 has dimension and support & E31 Document & xsd:string\\
    ISAD Ontology & --- ISAD7 has administrative history & E31 Document & xsd:string\\
    ISAD Ontology& --- ISAD8 has archival history & E31 Document & xsd:string\\
    ISAD Ontology & --- ISAD9 has scope & E31 Document & xsd:string\\
    ISAD Ontology & --- ISAD10 has access condition & E31 Document & xsd:string\\
    ISAD Ontology & --- ISAD11 has current quota & E31 Document & xsd:string\\
    ISAD Ontology & --- ISAD12 has old quota &	E31 Document & xsd:string\\
    ISAD Ontology & --- ISAD13 has original quota &	E31 Document & xsd:string\\
    ISAD Ontology & --- ISAD14 has language & E31 Document & xsd:string\\
    ISAD Ontology & --- ISAD15 has related unit of description & E31 Document & xsd:string\\
    ISAD Ontology & --- ISAD16 has existence and location of copies &	E31 Document & xsd:string\\
    ISAD Ontology & --- ISAD17 has publication notes & E31 Document & xsd:string\\
    ISAD Ontology & --- ISAD18 has notes & E31 Document & xsd:string\\
    ISAD Ontology & --- ISAD19 has system of arrangement & E31 Document & xsd:string\\
    ISAD Ontology & --- ISAD20 has physical characteristics & E31 Document & xsd:string\\
    ISAD Ontology & --- ISAD21 has description date & E31 Document & xsd:string\\
    ISAD Ontology & --- ISAD22 has last modification & E31 Document & xsd:string\\
    ISAD Ontology & --- ISAD23 has predominant date & E31 Document & xsd:string\\
  \bottomrule
\end{tabular}
\end{table}

\end{document}